\documentclass[aps,prc,floats,floatfix,preprint,superscriptaddress,nofootinbib]{revtex4} 
 
\usepackage{graphicx} 
\def\be{\begin{equation}} 
\def\ee{\end{equation}}

\begin{document}

\title{Nonadiabatic generation of coherent phonons} 
 
\author{Y. Shinohara} 
\affiliation{Graduate School of Pure and Applied Sciences, University of Tsukuba,  
Tsukuba 305-8571, Japan}  
\author{S. A. Sato}  
\affiliation{Graduate School of Pure and Applied Sciences, University of Tsukuba,  
Tsukuba 305-8571, Japan}  
\author{K. Yabana}  
\affiliation{Graduate School of Pure and Applied Sciences, University of Tsukuba,  
Tsukuba 305-8571, Japan}  
\affiliation{Center for Computational Sciences, University of Tsukuba,  
Tsukuba 305-8571, Japan}  
\author{J.-I. Iwata}  
\affiliation{School of Engineering, The University of Tokyo, Bunkyo-ku 113-8656, Japan}  
\author{T. Otobe} 
\affiliation{ 
Advanced Photon Research Center, Japan Atomic Energy Agency, Kizugawa, Kyoto 
619-0215, Japan} 
\author{G.F. Bertsch} 
\affiliation{Institute for Nuclear Theory and Dept. of Physics, 
University of Washingtion, Seattle, Washington}

\begin{abstract} 
The time-dependent density functional theory (TDDFT) is the leading  
computationally feasible theory to treat excitations by 
strong electromagnetic fields.  Here the theory is applied to 
coherent optical phonon generation produced by intense laser pulses. 
We examine the process in the crystalline semimetal antimony (Sb), 
where nonadiabatic coupling is very important.   
This material is of particular interest because it exhibits strong  
phonon coupling and optical phonons of different symmetries can be observed. 
The TDDFT is able to account for a number of qualitative features of  
the observed coherent phonons, despite its unsatisfactory performance 
on reproducing the observed dielectric functions of Sb.  A simple dielectric
model for nonadiabatic coherent phonon generation is also examined 
and compared with the TDDFT calculations. 
\end{abstract} 
 
\maketitle 
 
\section{Introduction} 
 
In this paper, we apply time-dependent density functional theory 
(TDDFT) to calculate coherent phonon generation in crystalline solids. 
There is fairly clear separation between adiabatic and nonadiabatic  
regimes for this process, depending on the the material and the frequency  
of the external electromagnetic field.  We first summarize the physical  
aspects of the phonon generation. 
 
\begin{figure} 
\includegraphics[scale=0.6] {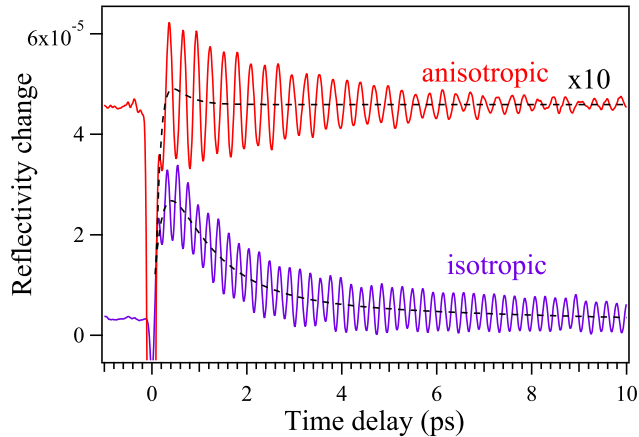} 
\caption{\label{ishioka} Observation of coherent phonons 
in crystalline Sb generated by high-intensity laser pulses of 
1.55 eV photon energy, from \cite[Fig. 2a)]{is08}. 
} 
\end{figure}

Coherent optical phonons generated by high-intensity, ultrashort laser pulses  
can be easily observed by pump-probe experiments that are sensitive 
to the changes in the index of refraction of the probed material. 
In particular, the phases of the phonons can be extracted from 
the reflectivity change plotted against the delay time of the reflected 
probe pulse.  
These experiments have been done for many kinds of materials.  
The coupling to optical phonons is especially strong in the  
semimetals Bi and Sb, and our calculations here will be for crystalline Sb. 
An example of the kind of data that motivates this choice are shown  
in Fig. \ref{ishioka} \cite{is08}. 
The pump and the probe pulses are directed  
on the $[0{\bar 1}12]$ surface of a Sb crystal at nearly normal incidence.   
The change in the reflectivity of the probe pulse is measured as a  
function of the delay between the pump and the probe.   
One sees an oscillatory pattern whose frequencies can be identified  
with the known optical phonons in the crystal.   
The crystal symmetry imposes some conditions between the polarization  
of the pump pulse and the probe pulse.  That information has been  
used in the experiment to separate the contribution of the $E_g$ phonon  
from that of the $A_{1g}$ phonon.  The signal label ``isotropic" is due to 
the $A_{1g}$ phonon while the one labeled ``anisotropic" is 
due to the $E_g$ phonon. 
  
For a given phonon, the reflectivity change is often parameterized 
by the functional form 
\be 
{\Delta R\over R} = g e^{-\Gamma t} \cos(\omega_{ph} t + \phi), 
\ee 
where $g$ is the amplitude, $\omega_{ph}$ is the phonon angular frequency,  
$\Gamma$ is a damping constant, and $\phi$ is a phase angle.  
The phase angle $\phi$ provides a very sensitive measure of the  
mechanism for the phonon generation.  If the mechanism is adiabatic,  
the phase angle should be close to $\pi/2$, as will be discussed below. 
Typically this is achieved in insulators when the laser photon 
energy is below the direct band gap.   
This is called the "impulsively stimulated Raman scattering" (ISRS) 
mechanism \cite{ya85}, 
because the entire process can be described in terms of the Raman  
couplings for exciting a single phonon by a single photon. 
This process occurs for laser pulses whose duration is shorter than 
the vibrational frequency.   
In this mechanism, the coherent phonon coupling depends only on measurable  
dielectric properties of the medium.  The equation of motion in the phonon 
coordinate is given by a simple formula of Merlin\cite{me97}, 
\be 
\label{ad_eq} 
\frac{d^2 q}{d t^2} + \omega_{ph}^2 q = F(t) = 
\frac{1}{2}\frac{\partial \chi}{ \partial q} | {\cal E}(t)|^2\,. 
\ee 
Here $q$ is the phonon coordinate, $\chi$ is a component of the dielectric  
susceptibility tensor, and $\cal E$ is a corresponding component of the  
electric field. 
 
When real excitations of the medium are possible, another process called 
the displacive mechanism can contribute as well \cite{ze92,sc93,ku94}. 
This mechanism takes place in opaque materials like semimetals and  
in insulators when the laser frequency is above the direct band gap. 
In the displacive mechanism, the electronic excitations produce a long-term 
shift in the charge distribution, changing the equilibrium position 
of the phonon coordinates.  Thus the pump pulse produces a state in which 
the phonon coordinates are displaced from their new equilibrium positions. 
If the life-time of the excited charges is sufficiently long, 
the oscillation of the phonons about the new equilibrium will be given 
by a cosine functional dependence on the time with respect to the  
pump pulse.  Thus, the displacive mechanism has a phase differing by 
$\pi/2$ with respect to the ISRS mechanism. 
 
At first sight it would seem that the ISRS and the displacive mechanisms 
are quite different physical processes. 
In an attempt to establish a unified description, 
Stevens, Kuhl, and Merlin (SKM) proposed a unified model of 
to describe both ISRS 
and displacive mechanisms in term of the dielectric properties of the medium \cite{st02}. 
Their approximate 
formula for the Fourier component of the force, $F(\Omega)$, 
is given by, 
\be 
\label{Fomega} 
F(\Omega) = C 
\left[ {d {\rm Re}(\varepsilon)\over d \omega} + \frac{2 i {\rm Im}(\varepsilon)} 
{\Omega} \right] 
\int_{-\infty}^{+\infty} e^{i\Omega t} \vert E(t) \vert^2 dt, 
\ee 
where $E(t)$ is the laser electric field, $\omega$ is the laser frequency,
and $\varepsilon$ is the dielectric function. 
In Sect. IV below we will compare the formula with the results of the TDDFT 
dynamics to assess the reliability of the approximations made in deriving
it.
  
The transition from adiabatic to nonadiabatic coupling  
has been observed in crystalline Si\cite{ha03,ri07,ka09} as 
a rather rapid change of phonon phase as the laser photon energy  
crosses the direct band gap.   
We have previously applied TDDFT to this system 
and found that it clearly reproduced this transition 
\cite{sh10,note1}.  In this work we will use the same computational 
framework, but applied to a semimetal rather than a semiconductor 
having a very well defined band gap.  However, due to the different 
crystal symmetry (A7 rather than cubic) the codes had to be  
significantly modified to treat Sb.  In the next section we briefly  
summarize the computational aspects, in particular the extensions  
needed for the present application.   
 
\section{Time-dependent density functional theory} 
\label{TDDFT} 
 
We have found that the Lagrangian formulation of the dynamics problem 
is very helpful not only from a formal point of view but also to 
construct the computational equations of motion satisfying the  
necessary conservation laws.  The Lagrangian we used in our earlier 
study \cite{sh10,diamond}  contains the following elements:\\ 
1)  a fully microscopic treatment for the electron dynamics  
using a Kohn-Sham (KS) energy functional to evolve the time-dependent  
electron orbitals;\\ 
2)  a classical treatment of the time-dependent electric  
field in the crystalline unit cell;\\ 
3)  a classical treatment of the dynamics of ionic centers, 
often called ``Ehrenfest dynamics". 
 
We write the Lagrangian as a sum of three terms, the Kohn-Sham, 
electromagnetic, and ionic parts. 
\be 
L =  L_{KS} + L_{em}+L_{ion}. 
\ee 
The Kohn-Sham term is given the following integral over the unit
cell $\Omega$,
\be 
L_{KS} = \sum_i \int_{\Omega} d\vec r 
\left\{ \psi_i^* i\frac{\partial}{\partial t} \psi_i 
-\frac{1}{2m} \left\vert \left( -i\vec\nabla 
+ \frac{e}{c} \vec A \right) \psi_i \right\vert^2 \right\} 
-\int_{\Omega} d\vec r \left\{ \left( en_{ion} - en_e \right) \phi 
-E_{xc}[n_e] \right\}. 
\ee 
The variables here are the electron orbitals $\psi_i({\vec r},t)$, 
the electric field potentials, $\phi({\vec r},t)$ and 
$\vec A(t)$, and the ionic coordinates $\vec R_\alpha(t)$. 
The vector potential ${\vec A}(t)$ is a function of time 
without spatial dependence and describes spatially-uniform  
electric field, while a scalar potential $\phi({\vec r},t)$  
is periodic in the unit cell.  
The separation of the electric field into these two components
is crucial to our computational scheme \cite{biry,ot08}.  
It enables us to apply Bloch theorem for 
electron orbitals $\psi_i$ at each time. The ionic density 
$n_{ion}$ is described with the ionic coordinates $R_\alpha$ 
and the electron density $n_e$ with the Kohn-Sham orbitals. 
 
We employ the same exchange-correlation energy  
functional $E_{xc}[n]$ for dynamical calculation as that 
is used for the ground state calculation. This is the well-known  
``adiabatic approximation" in time-dependent 
density functional theory.  Of course, the electron  
dynamics in an external field can be highly non-adiabatic. 
 
The electromagnetic Lagrangian is taken as 
\be 
L_{em}=\frac{1}{8\pi} \int_{\Omega} d\vec r 
\left\vert \vec \nabla \phi \right\vert^2 
+ \frac{\Omega}{8\pi c^2} \left( \frac{d\vec A}{dt} \right)^2. 
\ee 
This form is sufficient to treat the coupling in the medium 
at length scales small compared to the photon wave length. 
For the full electrodynamics including the transmission and 
reflection of photons from the crystal surface, the Lagrangian
must also include magnetic fields.
This has been carried out in another context, namely the 
deposition of energy by strong laser pulses\cite{ya12}. 
 
We separate the vector potential $\vec A(t)$ into  
external field contribution $\vec A_{ext}(t)$ and induced  
polarization $\vec A_{ind}(t)$. Whether to include the  
induced polarization or not in $\vec A(t)$  
depends on the macroscopic geometry of the sample 
and the polarization direction of the electric field. 
In the present calculation, we employ the longitudinal  
geometry in which the induced field is included in $\vec A(t)$. 
 
Finally, the dynamics in the ion coordinates $\vec R_\alpha$ is 
governed by the classical Lagrangian, 
\be 
L_{ion} = \frac{1}{2}\sum_{\alpha} M_{\alpha} \left( \frac{d\vec R_{\alpha}}{dt} \right)^2 
+\frac{1}{c}\sum_\alpha Z_\alpha e \frac{d \vec R_\alpha}{dt}\vec A. 
\ee  
 
At a formal level, we were able to prove that the TDDFT dynamics 
reduces to the ISRS formula Eq. (2) in the limit where the 
laser pulse does not deposit energy into the electronic degrees 
of freedom.  In this adiabatic regime, the relevant 
dielectric properties can be calculated in perturbation theory 
based on static density functional theory (DFT), as was done in an early  
calculation of Raman scattering in Si crystals \cite{ba86}.   
In our full TDDFT calculation in Ref. \cite{sh10},  we found that  
the theory could describe both the adiabatic ISRS and the displacive 
mechanisms of excitations, thus providing a comprehensive framework  
for treating coherent laser-lattice interactions. 
 
\section{Application to Antimony} 
 
\subsection{Numerical implementation} 
 
The solver for the time-dependent Kohn-Sham (KS) equation  
is a key element for practical computations.   
Our implementation of the KS solver uses a 3-dimensional 
real-space grid to represent the orbital wave functions \cite{ya96}.   
This is straightforward for molecules and other finite systems as 
well as extended materials with cubic crystalline symmetry 
such as Si.  However, Sb has only a rhombic crystal symmetry,  
and the grid must be modified accordingly.   
Fortunately, crystals such as the rhombic have a 3-fold  
symmetry axis, allowing the primitive hexagonal unit cell to  
be embedded in a supercell having the shape of  
a rectangular parallelepiped.  The construction is shown in  
Fig. 2. The hexagonal face on the top of the cell is replaced 
by a rectangle of axes ratio $\sqrt{3}:1$.  The Cartesian 
periodicity needed by the KS solver can then be achieved by 
changing the mesh spacings in each dimension so that the 
supercell is spanned by an integer number of mesh points in 
each direction.   
 
In the case of Sb, the crystal structure is only slightly distorted 
from cubic, which helps one to construct the lattice as well as 
to understand the character of the optical phonons.  The idealized 
undistorted lattice is constructed from a simple cubic lattice as 
shown in Fig. 2.  This structure can also be viewed as two  
interpenetrating face-centered cubic lattices, distinguished in 
the figure by the red and blue colors of the atoms. 
The $c$-axis of the cell goes along the $[111]$ direction of the cubes.   
The length of the $c$-axis in the undistorted geometry is $\sqrt{6}$  
larger than the short axis.  The actual structure is now obtained by  
making two transformations. First, the $c$-axis is elongated by  
a factor of 1.065.  Second, one of the cubes is displaced by a small  
amount (3.3\%) along the $c$-axis with respect to the middle of  
the other cubic.  This displacement of the two sublattices with  
respect to each other define the coordinates of the optical phonons.   
Displacements along the $c$-axis give rise to the $A_{1g}$ phonon.   
Displacements perpendicular to the $c$-axis give rise to the  
doubly-degenerate $E_g$ phonon.  In a perfectly cubic system  
all three modes would be degenerate.   
The offset equilibrium position in the distorted 
lattice breaks the degeneracy between the frequencies of the 
$A_{1g}$ and $E_g$ phonons, 4.65 THz and 3.47 THz respectively. 
 
For our representation of Sb, the supercell contains 12 atoms  
and has dimensions 
$(1,\sqrt{3},\sqrt{6}*1.065)a$ with $a=8.12$ au. 
We take a mesh of $14 \times 30\times 48$ points giving mesh  
spacings of $(0.58 ,0.47, 0.44)$ au.   
The solver is described in detail in previous publications.   
Here we just note that we use a time step of $\Delta t = 0.04$ au,  
which is sufficiently small for the time evolution to be stable. 
Typically electron orbitals are evolved for 20,000 time steps. 
 
At each time step, the KS wave functions are calculated for a 
grid of $16^3$ $k$-points in the Brillouin zone.
The densities for each grid point are summed to  
obtain an updated KS operator for the next time step.   
It is convenient to parallelize the code by distributing the  
calculations for the $k$-points to different processors. 
 
\begin{figure} 
\includegraphics[scale=1.0] {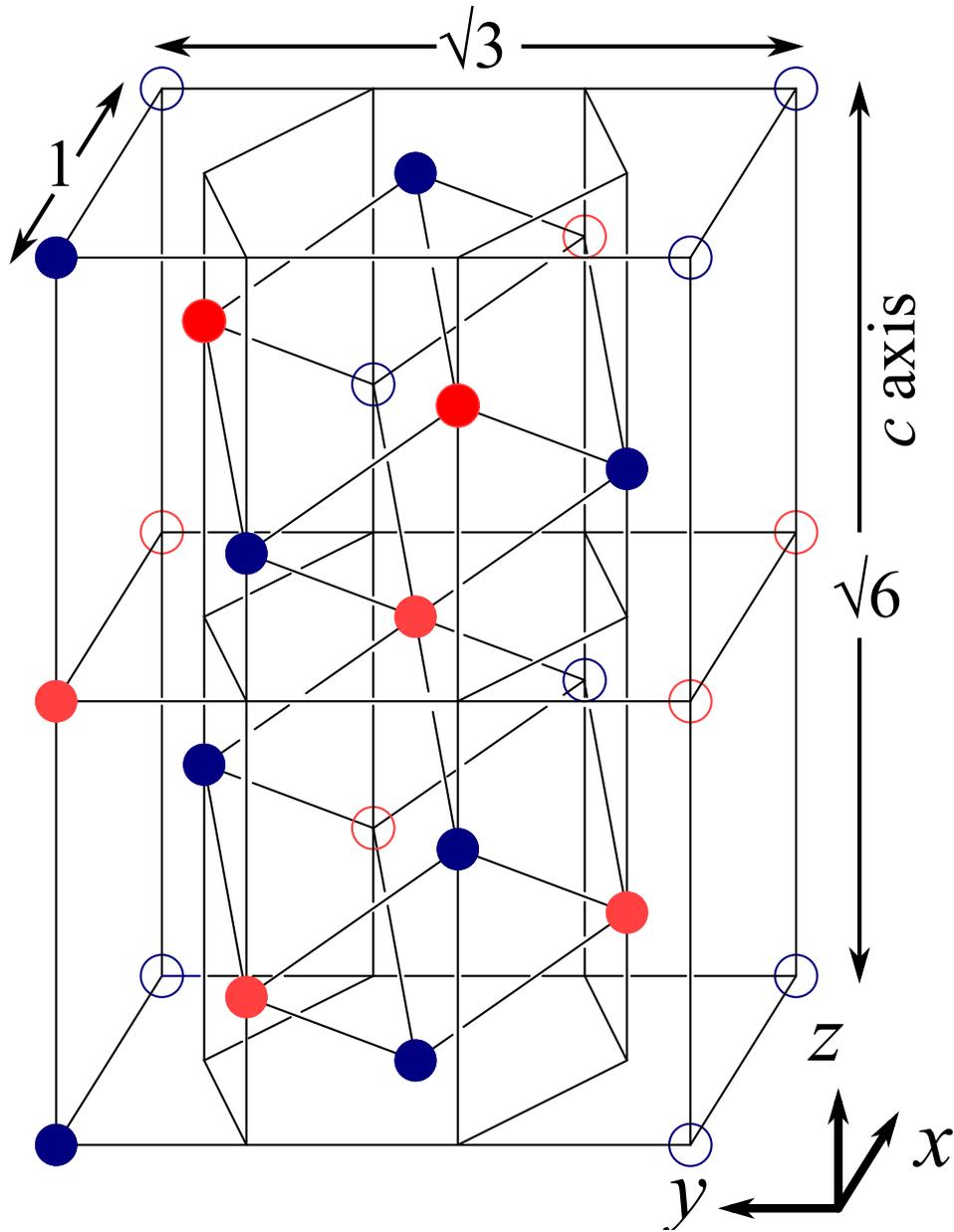} 
\caption{\label{orthorhombic} Idealized cubic unit cell illustrating 
the embedding of the orthorhombic unit cell onto the rectangular 
supercell.  The orthorhombic cell is outlined by the hexagonal  
prism.  Two simple cubes are shown with their $[111]$ axes along the $c$-axis 
of the supercell.  The 12 atoms of the supercell are shown with 
filled circles.  The red and blue atoms define interpenetrating 
face-centered lattices.  The $c$-axis is slightly enlongated to obtain 
the actual rhombic lattice.  In addition, there is a small shift of the  
red with respect to the blue sublattice along the $c$-axis. 
} 
\end{figure} 
 
The ionic motion is very small during the time of passage of 
the pump pulse, and 
we don't attempt to solve Newton's equations directly within the 
time-dependent evolution of the system.  As in Ref. \cite{sh10}, 
the code only calculates the force on the ions.  This is later
decomposed 
a transient part and a constant part in the final state for the
final analysis. 
 
The energy functional in our calculations treats explicitly the 
5 electrons in the $(5s)^2(5p)^3$ atomic orbitals.  The interaction 
with the ionic cores is taken as the Troullier-Martins \cite{tr91}  
pseudopotential with the range of the nonlocality of 4.0 au. 
The electron-electron interaction is treated in the local density  
approximation using the standard parameterization \cite{pz81}.   
 
\subsection{Calculated electronic structure} 
 
The  group V elements, As, Sb, and Bi, have a complicated band 
structure due their rhombohedral A7 crystalline symmetry.   
They are distinguished by a small but nonzero overlap between  
nominally valence and conduction bands.   
This gives the crystal the characteristics of a semimetal,  
namely a conductor with a very small carrier density. 
There are several DFT calculations of the electronic structure  
which reproduce this important feature\cite{go90,sh99}.  
For example, the calculation of Ref. \cite{go90} found  
a carrier density of $2.3\times10^{-3}$  electrons per atom in Sb  
compared to the experimental value of $1.1\times10^{-3}$.   
Our DFT  calculation gives a similar density of  
electron states and a carrier density of $2.4\times10^{-3}$  
electrons per atom.  In any case, details of the density of states  
within a few tenths of eV of the Fermi level should not be  
crucial to the dynamics at the much higher energy of the  
laser photon.  
 
Beyond the static electronic structure, it is important to  
establish the accuracy of the linear response predictions  
if the dynamic electronic properties are the object of the  
calculations. 
There do not seem to be any DFT calculations of the dielectric  
properties in the literature, so we describe our results in some  
detail in the Appendix.  Unfortunately, our predicted dielectric  
properties do not agree well with the evaluated measurements  
\cite{pa98}.  However, the evaluated dielectric function is  
derived from the measured reflectivity function which seems to 
depend significantly on temperature.  In fact our predicted  
reflectivity agrees fairly well with the data at 77K but not with  
room temperature data.  Since that experimental finding seems  
not to be understood up to now, we cannot draw any strong  
conclusions about the origin of the disagreement with theory. 
In any case, our TDDFT results can be compared with the  
simple models to evaluate the reliability of the approximations  
made in the models. 
 
\subsection{Typical results} 
 
In this section, we show the results of the time-dependent calculations 
for typical conditions.  As shown in Fig. 2, we set the $z$-axis  
in our calculation parallel to the $c$-axis. The form of the external 
electric field is chosen as 
\be 
\vec{\cal E}_{ext}(t) = \hat x {\cal E}_0\sin(\pi t/T)\sin(\omega t), 
\ee 
where the polarization direction is chosen along the $x$-axis, 
${\cal E}_0$ is the maximum value of the electric field, 
$\omega$ is the laser frequency, and $T$ is the pulse duration.  
Crystal symmetry permits this field to excite the $A_{1g}$ mode 
and the $x$ component of  the $E_g$ mode.   
 
\begin{figure} 
\includegraphics[scale=1.0] {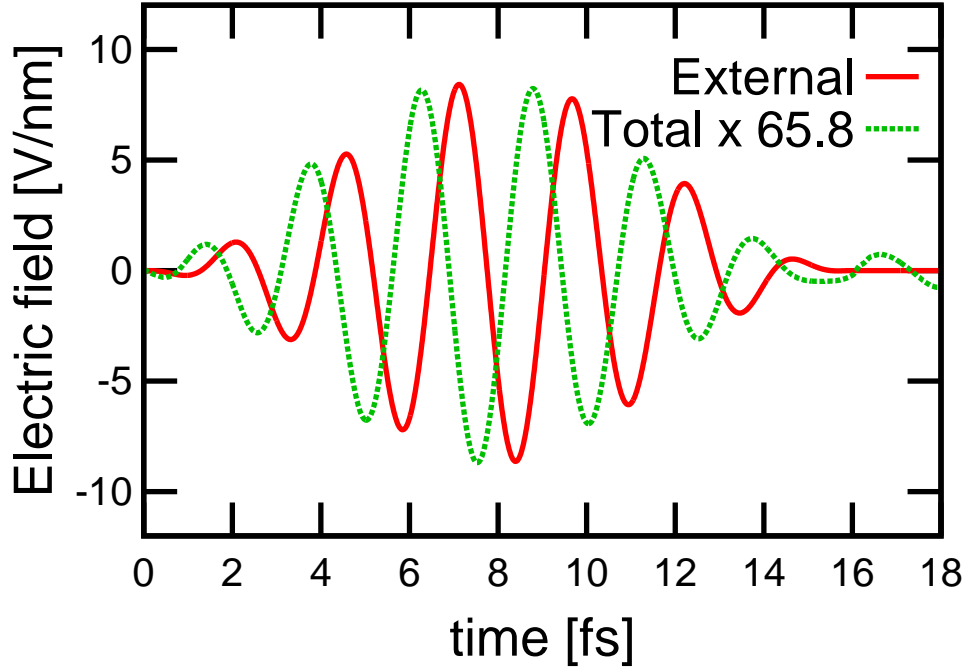} 
\caption{\label{x-EF} External and screened electric fields in 
the Sb crystal.  See text for explanation. 
} 
\end{figure}

As we mentioned in Sec. \ref{TDDFT}, we include the induced  
polarization field 
in the vector potential $\vec A(t)$. For a sufficiently weak electric 
field, the external and the total (external + induced) electric fields  
are related to each other by the dielectric function, 
\be 
{\cal E}_{tot}(t) = \int^t \varepsilon^{-1}(t-t') {\cal E}_{ext}(t') dt'. 
\ee 
The external and total fields for a typical run are shown in Fig. \ref{x-EF}.  
Here the parameters for the external field are taken as
$\hbar \omega= 1.6$ eV, $T=16$ fs, and  
${\cal E}_0$ corresponding to an intensity of $10^{13}$ W/cm$^2$.   
 
We see that the total field is almost out of phase to the external field  
and about 66 times smaller in amplitude.  The ratio and the phase  
between the two corresponds well to the calculated dielectric function  
at $\omega = 1.6$ eV/${\hbar}$, 
shown in Fig. \ref{eps_r_i}, even though the dielectric function is  
only strictly valid  in the small amplitude regime. 
 
\begin{figure} 
\includegraphics[scale=0.6] {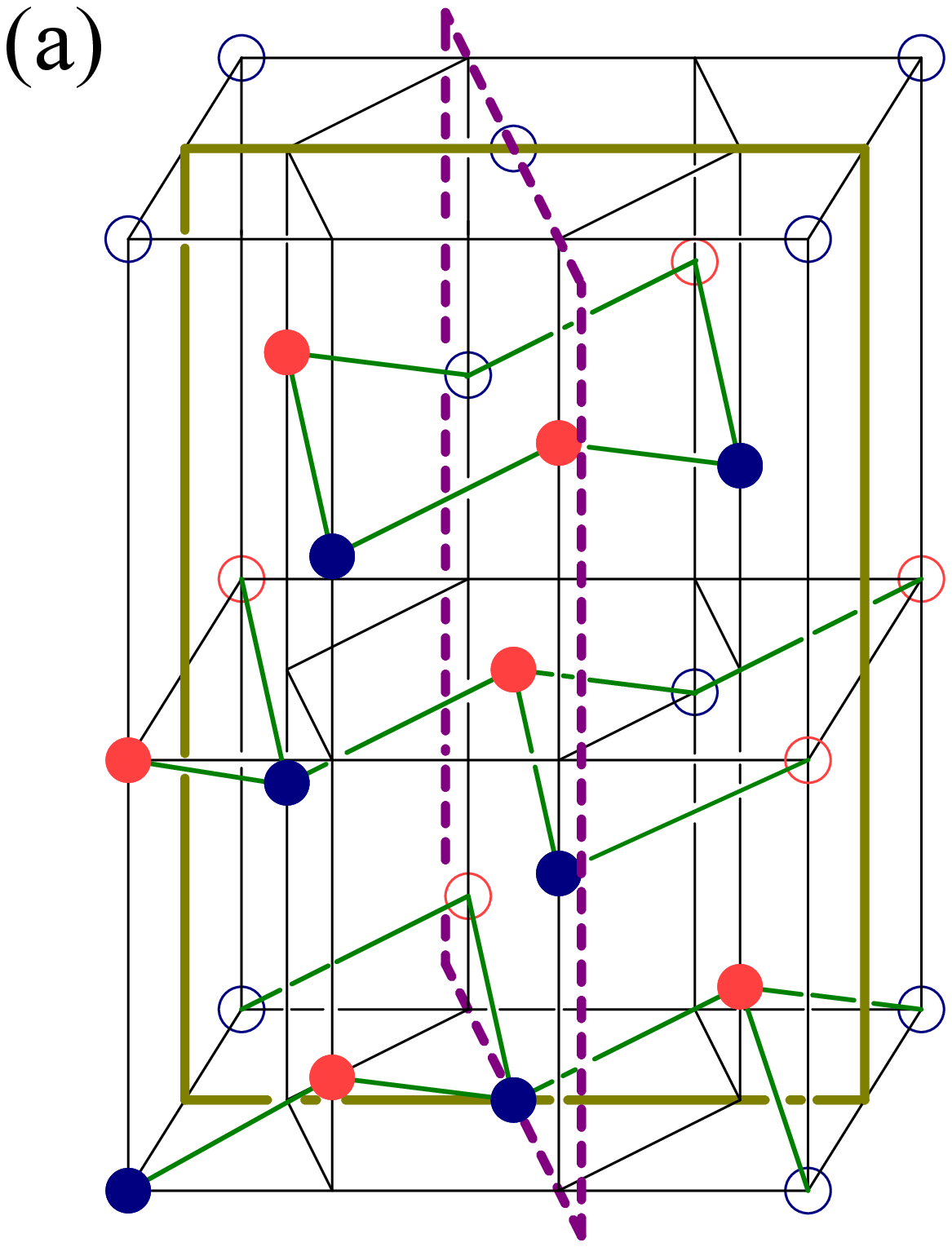} 
\includegraphics[scale=0.6] {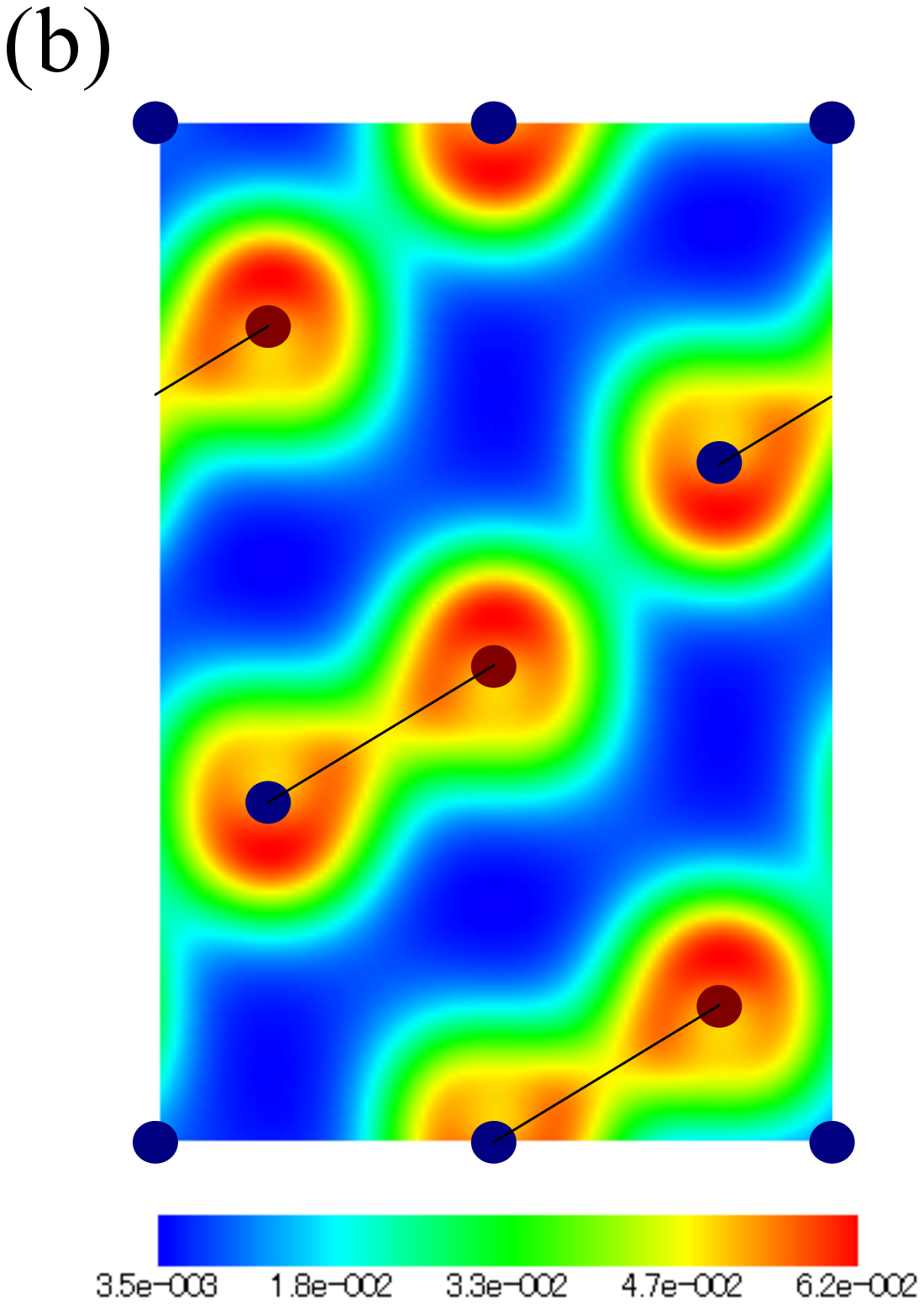} 
\includegraphics[scale=0.6] {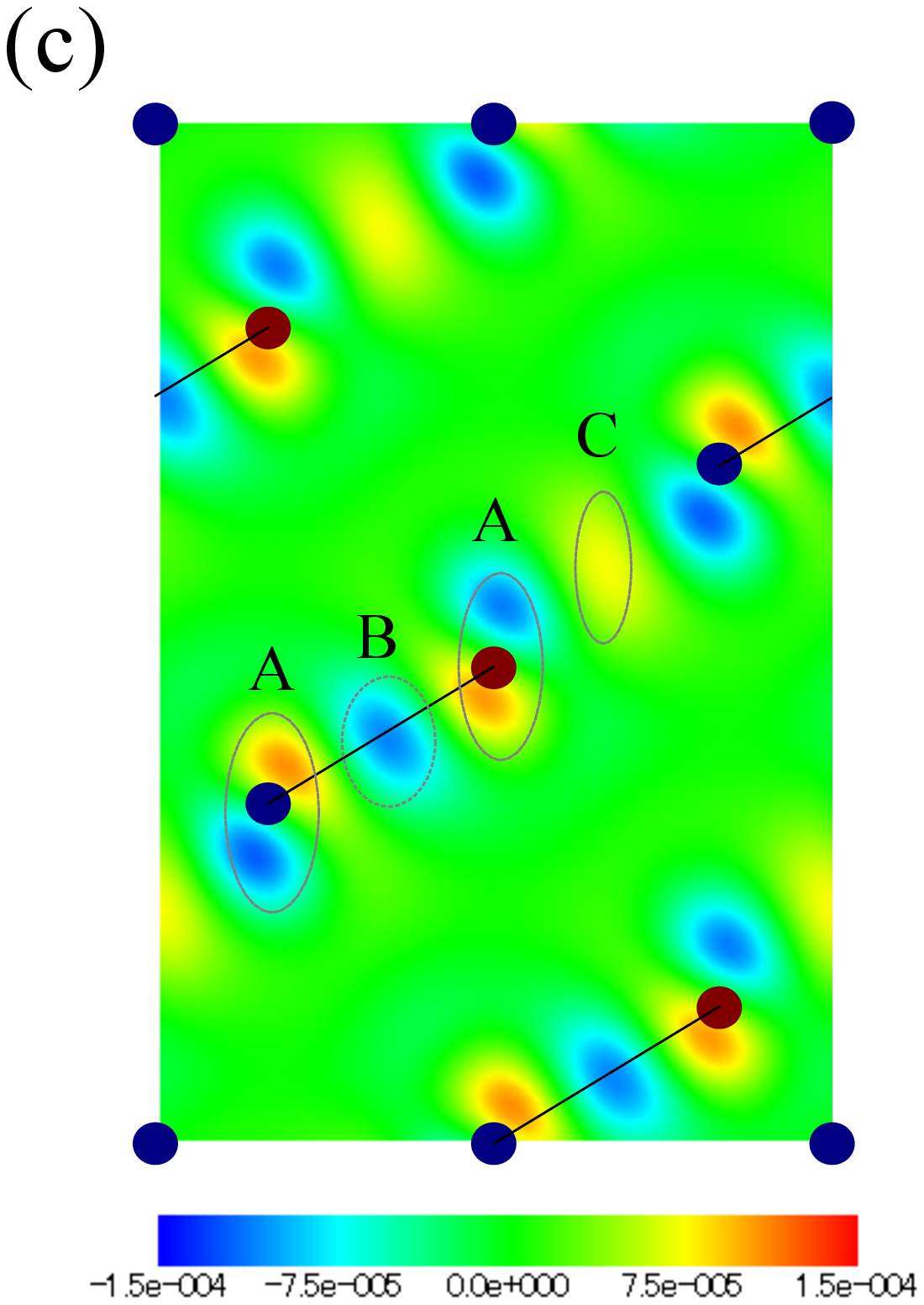} 
\includegraphics[scale=0.6] {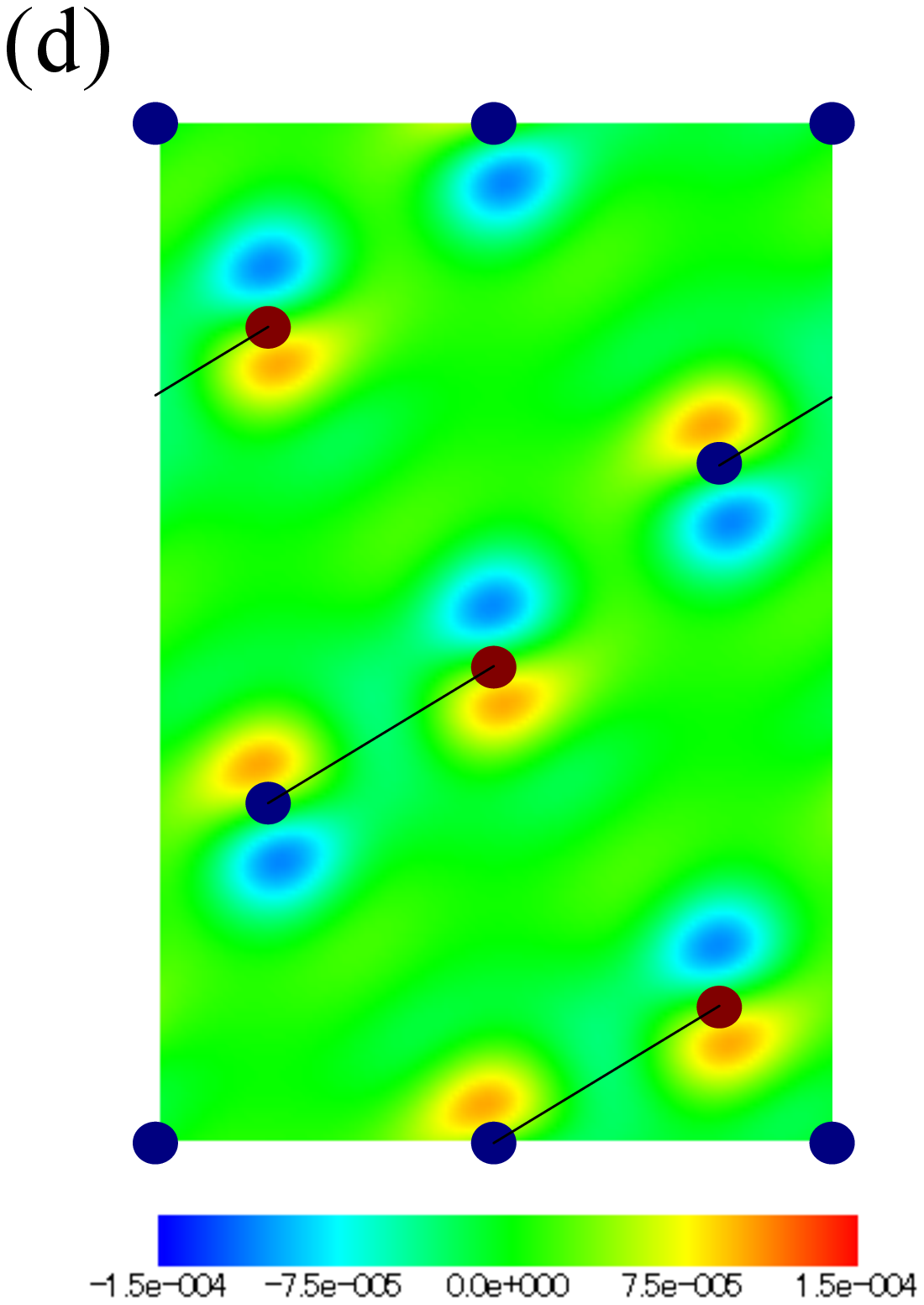} 
\caption{\label{density} (a) Atomic positions in the rectangular 
unit cell. An $xz$-plane through the middle of the cell is framed
in a solid green rectangle.
The purple-dashed frame marks the plane obtained by rotating  
green-solid frame by 120$^{\circ}$ around the $c$-axis.  
(b) The ground state electron density on the two planes.
(c) Density change in the final state on the $xz$ plane
(green-solid frame).
(d) Density change on the rotated plane (purple-dashed frame). 
} 
\end{figure}

The electron density in two planes through the unit cell are shown in
Fig. \ref{density}.
In panel (a), the atomic positions in the rectangular  
supercell are depicted with the actual distortion of the cubic lattice into
rhombohedral.
The electron densities are shown  
in panels (b)-(d) on planes indicated by green-solid and purple-dashed  
frames. The green-solid frame is the $xz$-plane through the middle  
of the unit cell. This plane includes the polarization direction of  
the electric field as well as the $c$-axis of the crystal.   
The purple-dashed frame is obtained by rotating the $xz$-plane  
by 120$^{\circ}$ around the $c$-axis passing through the central atom. 
 
The ground-state electron density, shown in (b), is the same in the
two planes.  
Each atom has three bonds with nearest neighbors and one of the bonds 
lies within each plane. 
Among five valence electrons, three of them are associated with
the bonds and two of them occupy lone-pair orbitals. 
 
The particle-hole excitation changes the occupation probabilities in
the final state, affecting the electron density distribution.
This is shown in  
two panels, (c) and (d). The panel (c) shows the change  
of electron density from that in the ground state in the  
$zx$ plane of the green-solid frame in (a). The orange color  
indicates increase of the electron density from the grand state, 
while the blue color indicates decrease. The panel (d) 
shows the change of electron density in the plane of 
purple-dashed frame in (a). 
 
There are two spatial regions where electron  
density changes most from that in the ground state. One is the  
excitation from lone-pair orbitals. This corresponds to 
the region just above and below the positions of atoms  
indicated as A in (c). This change is seen both  
in (c) and (d). The other is the excitation out of the bonding orbitals, 
causing a decrease of the density around the midpoint of the bonds. 
This removal  
of bonding electrons is seen clearly in (c) 
in the areas indicated as B in (c).  
The effect is much smaller in (d) because the electric field vector
is not in the plane of the that bond.
We also note that increase of electrons density is seen  
in the area C in (c), not in (d). 
These anisotropic changes of electron density certainly contribute
to the force on the optical phonon modes, which we now discuss.
 
\begin{figure} 
\includegraphics[scale=0.5] {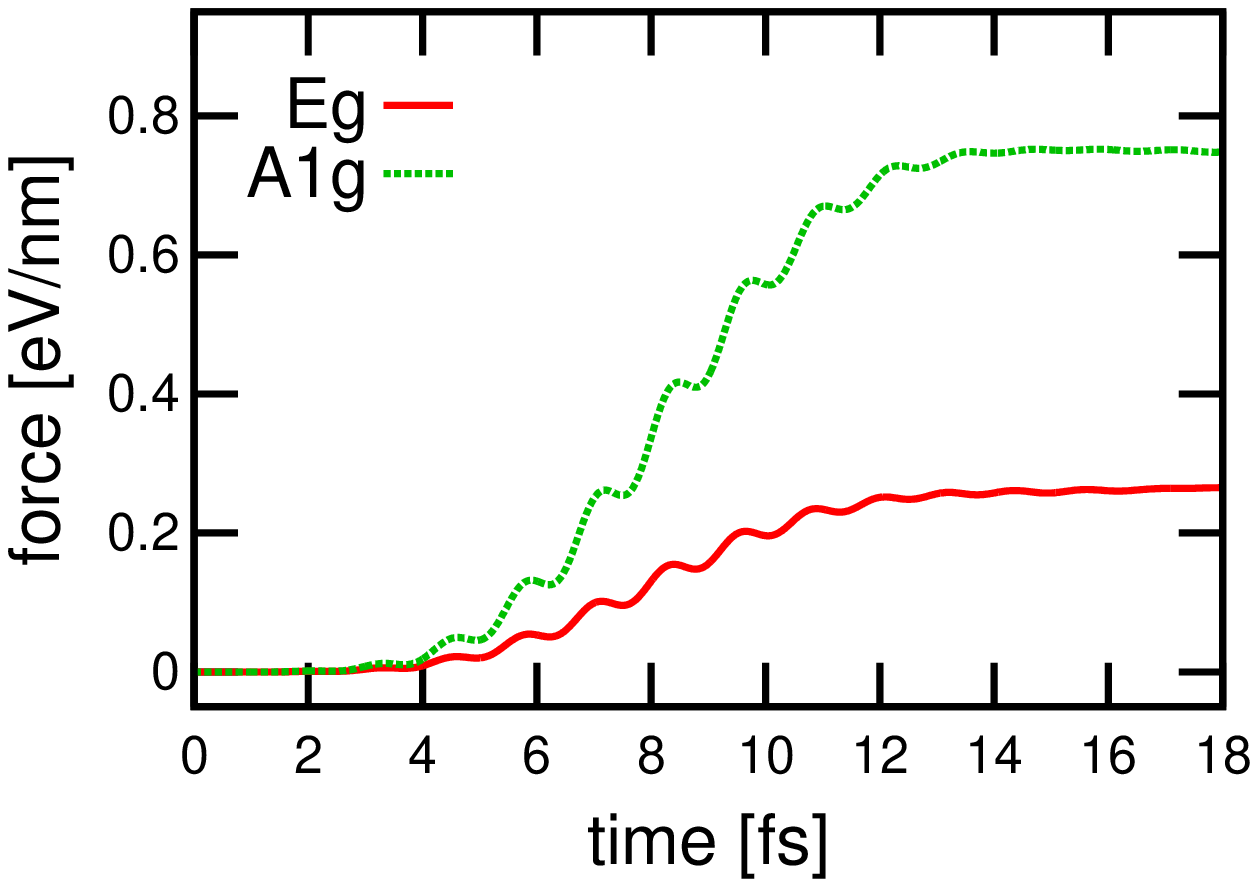} 
\includegraphics[scale=0.5] {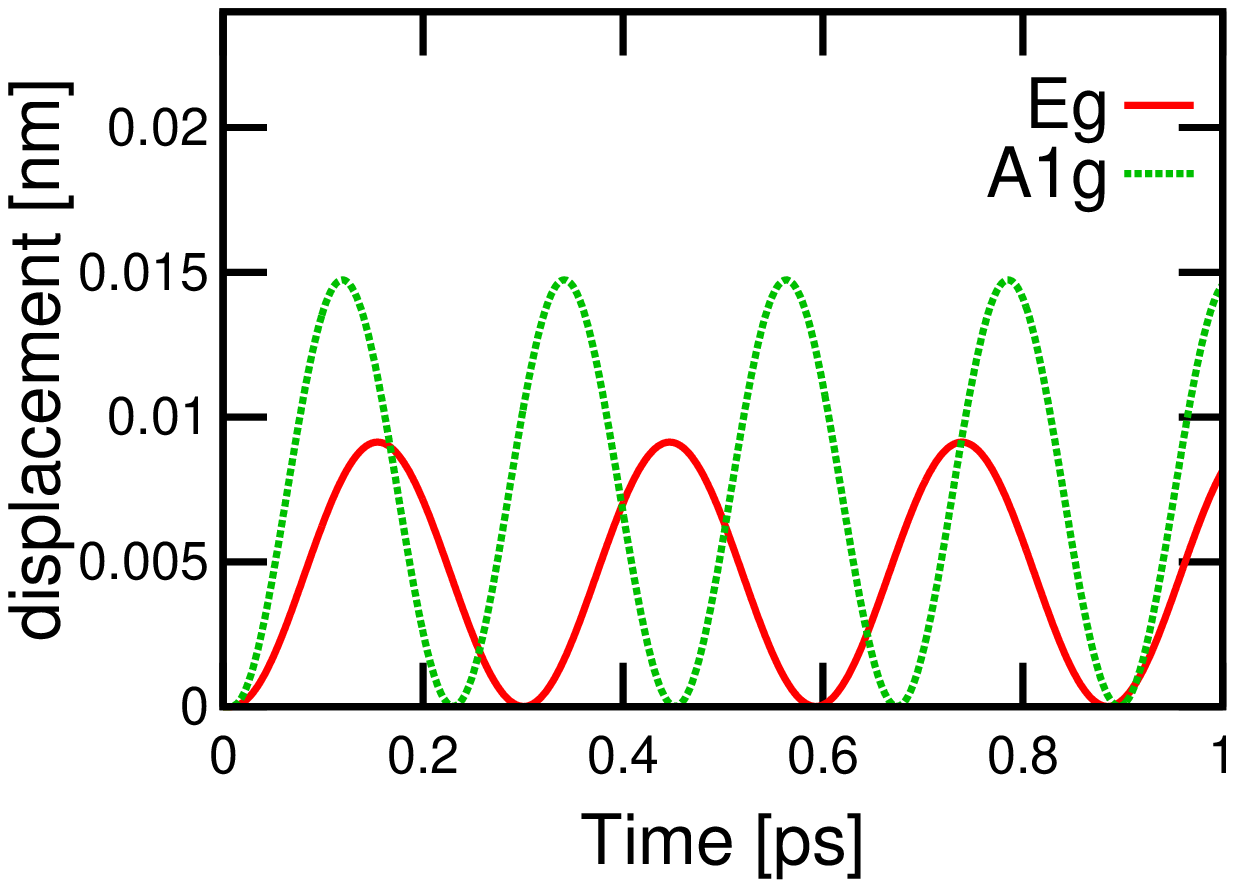} 
\caption{\label{force-x}  a) Force on phonon modes excited by the  
external field in the $x$-direction. b) corresponding displacement 
of phonon coordinates. 
} 
\end{figure}

\begin{figure} 
\includegraphics[scale=0.5] {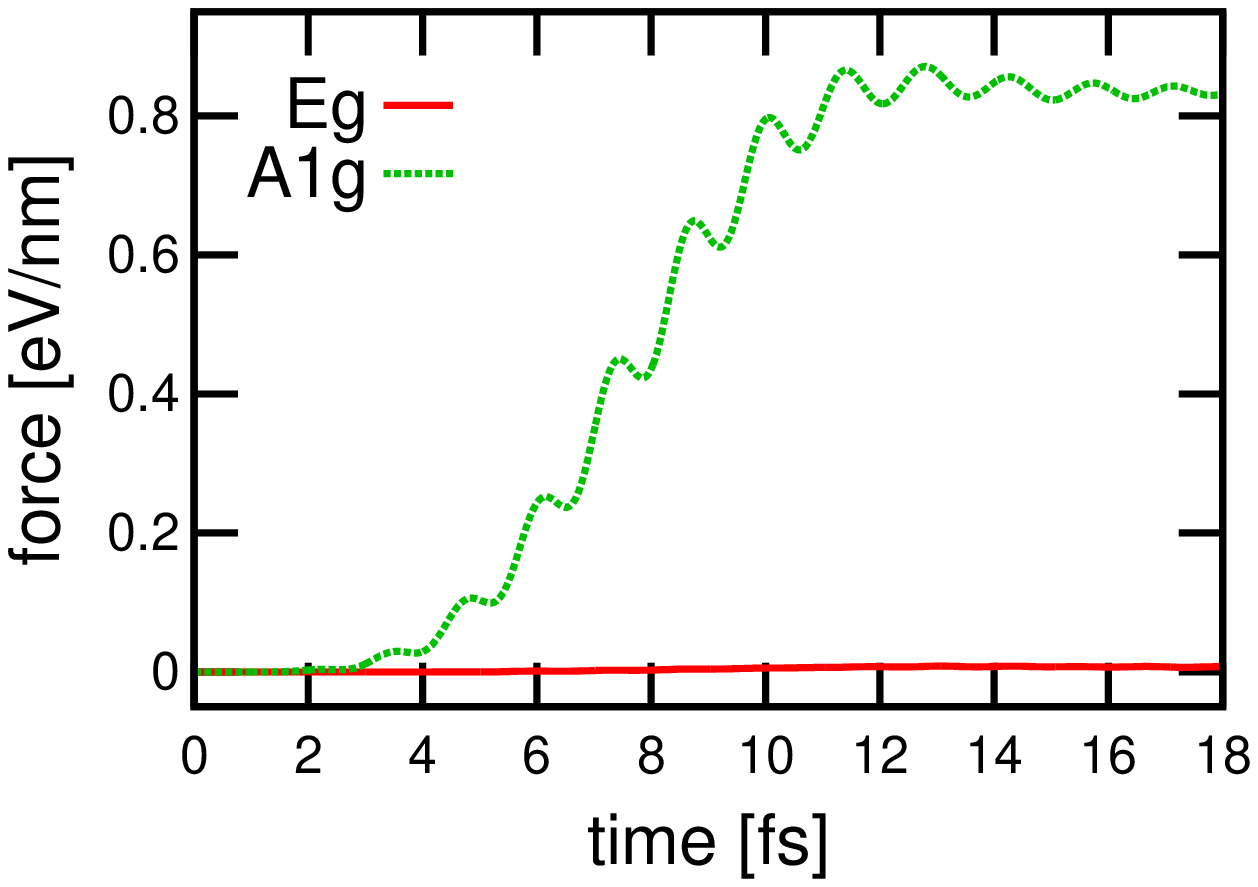} 
\caption{\label{force-z}  Force on phonon modes 
excited by the external field in the $z$-direction. 
} 
\end{figure}

The next two figures show the force on the phonon modes for the  
external field of Fig. \ref{x-EF}.  In Fig. \ref{force-x}, 
the external field is in the $x$ direction.  The $E_g$ phonon 
excitation will then also be in the $x$ direction; the $A_{1g}$ phonon 
is always along the $z$ axis. 
The oscillatory part of the force may be associated with the  
ISRS excitation mechanism.  The force reaches a constant plateau 
at the end of the pulse; its plateau value controls the strength 
of the displacive mechanism.  Note that the $A_{1g}$ mode is 
much stronger than the $E_g$ mode.  This agrees with the 
experimental measurements, which always see a larger effect for 
the $a_{1g}$ mode.
Fig. \ref{force-z} shows the forces resulting from the same  
external field but oriented along the $z$ axis.  The strength 
of the $A_{1g}$ force is about the same as in the other orientation,  
but now the $E_g$ force vanishes.  This is consequence of the crystal 
symmetry.  
In panel b) of Fig. \ref{force-x}, we integrate the equation of motion for 
the phonon mode to find the displacement as a function of time.   
Since the force is constant in the final state, the displacement function 
will be close to the form $1-\cos(\omega_{ph} (t-t_0))$ where $t_0$ is 
near the maximum amplitude point of the driving field.  Thus the 
phonon will have a displacive phase. In fact we find that the displacive  
mechanism dominates for all external field frequencies in the range  
of 1.0-3.0 eV/$\hbar$. 
 
We now turn to the phase of the coherent phonon.  The displacive 
mechanism dominates for both modes in the TDDFT calculation, as is
clear from the force plateaus in Fig. 5 and 6. 
The experiments reported in Refs. \cite{ga96, is08} obtain a phase consistent
with the displacive mechanism for the $A_{1g}$, in agreement with the TDDFT.
However, they find a difference in phase for the $E_g$ mode, which is
not explained by the theory.  One possible origin of the
phase differences could be differences in relaxation times of excited carriers 
responsible for two mode. Unfortunately, the electronic relaxation in
the final state is beyond the scope of the TDDFT.
 
\section{Comparison with the SKM model} 
 
In Ref. \cite{st02}, Stevens, Kuhl, and Merlin proposed a dielectric 
model for the force acting on phonon mode. Taking Fourier 
transform of Eq.~(\ref{Fomega}), we may obtain the force  
as a function of time, 
\be 
\label{Ft} 
F(t) = C \left[ \frac{d{\rm Re} (\varepsilon)}{d\omega} |E(t)|^2 
+ 2 {\rm Im} (\varepsilon) \int^t_{-\infty} dt' |E(t')|^2 \right]. 
\ee 
This formula indicates that the real part of $\varepsilon$  
is related to the impulsive force, $F(t) \sim |E(t)|^2$, corresponding  
to the ISRS mechanism, while the imaginary part of  
$\varepsilon$ gives a constant force at $t \rightarrow +\infty$  
corresponding to the displacive mechanism.  
In this section, we will make a theory-to-theory comparison of  
this equation within the TDDFT dynamics. We compare the 
force at $t \rightarrow +\infty$ as a function of laser frequency 
with the imaginary part of the dielectric function, and examine  
the validity of the dielectric model. 
 
\begin{figure} 
\includegraphics[scale=1.0] {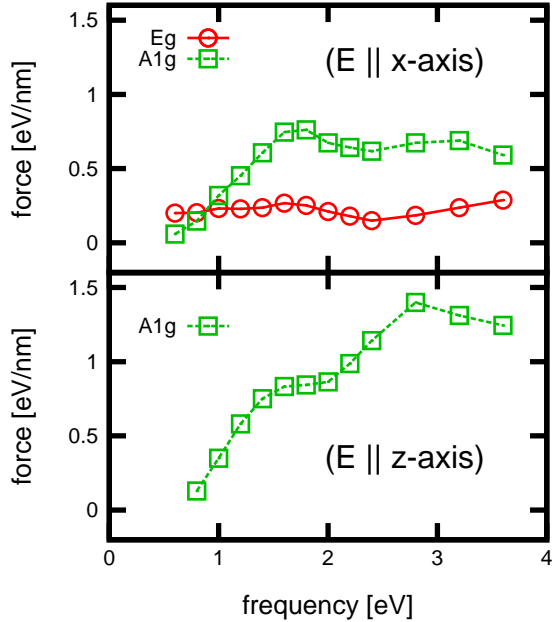} 
\caption{\label{f-vs-omega} Forces in the final state as a function 
of the field frequency.  
} 
\end{figure}

Figure \ref{f-vs-omega} show the plateau values of the force 
at $t \rightarrow +\infty$. In this figure, the intensity 
of the external electric field is taken to be the same. However,  
the electric field in Eq. (\ref{Ft}) is the actual electric field 
in the medium. 
We take this into account by scaling our calculated force
by $\vert\varepsilon(\omega)\vert^2$ to compare with 
the imaginary part of the dielectric function. 
 
\begin{figure} 
\includegraphics[scale=1.0] {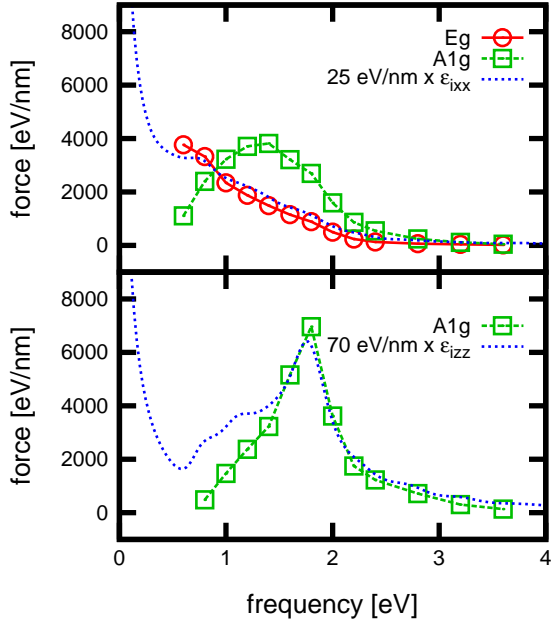} 
\caption{\label{skm} Comparison of forces calculated by  
the TDDFT compared to the SKM  model.  a) electric field in the 
$x$ direction; b) electric field in the $z$ direction. 
} 
\end{figure}

In Fig. \ref{skm}, we show the forces multiplied with  
$\vert\varepsilon(\omega)\vert^2$. The top panel shows  
the results for the electric field in the $x$-direction. 
The force calculated in TDDFT is shown as green squares  
and red circles. The bottom panel shows the quantities for  
the electric field in the $z$-direction. The $E_g$ force  
vanishes by symmetry and is not shown. We find the forces  
as a function of frequency exhibit different behavior depending 
on phonon modes and polarization of electric field. 
 
We test the SKM model by comparing the forces with  
the imaginary part of the dielectric function which is  
shown in the Appendix. 
In the top panel, the imaginary part of the dielectric function  
in $x$-direction, ${\rm Im} \varepsilon_{xx}$, normalized to  
reproduce the magnitude of the $E_g$ force (red circles) is shown. 
One sees that the model reproduces the frequency dependence  
of the $E_g$ force quite well. In the bottom panel,  
${\rm Im} \varepsilon_{zz}$ normalized to reproduce the magnitude  
of the $A_{1g}$ force (green squares) is shown. 
The model again reproduces the trend of the $A_{1g}$ force quite well. 
In both cases, the model describes the variation in a force  
when the phonon coordinates are parallel to the electric 
field. However, for the $A_{1g}$ force in the top panel, the force 
along $z$-direction in the $x$ polarization direction, the 
frequency dependence does not resemble either  
${\rm Im} \varepsilon_{xx}$ or ${\rm Im} \varepsilon_{zz}$. 
 
\section{Conclusions and outlook} 
 
The TDDFT has given us a calculational framework to study the transition 
from adiabatic to nonadiabatic processes in extended systems.  In  
previous work, we found that an important qualitative aspects of the 
transition, namely the phase of the coherently generated phonon, was 
correctly reproduced by the theory in crystalline silicon.  In this work, 
we have applied the same methodology to a more challenging material, 
namely the semimetal antimony.  The noncubic symmetry presents some 
computational problems that have been overcome.  The optical phonons 
have a more rich spectroscopy than in the cubic systems, and there 
are symmetry-dictated dependencies that are reproduced by the theory. 
Unfortunately, the present theory is not accurate enough at the linear 
response level to give quantitative predictions for the coherent phonon 
generation. At a qualitative level, the theory predicts a nonadiabatic 
response over a wide range of frequencies. At the experimentally  
measured frequency, the $A_{1g}$ mode shows the nonadiabatic behavior  
but the observed phase of the $E_g$ mode is different. 
 
This study points out the need for a more accurate theory of the  
electronic structure of Sb.  It might be that the $sp$ space is too 
restrictive; there is a closed $d$ shell close to the Fermi energy that 
could be easily polarized.  Also, the theory needs to be developed 
to treating the transmission and reflection of the electromagnetic 
pulses from the interfaces of the media, in order to describe  
the measurements quantitatively. 
 
\section*{Acknowledgment} 

The numerical results are obtained by early access to the K computer 
at the RIKEN Advanced Institute for Computational Sciences,
SR16000 at YITP in Kyoto University,
the supercomputer at the Institute of Solid State Physics, University
of Tokyo, and T2K-Tsukuba, University of Tsukuba.
This work was supported by the Strategic Programs for Innovative
Research (SPIRE), MEXT, and the Computational Materials Science
Initiative (CMSI), Japan, and by the Grant-in-Aid for Scientific Research,
MEXT, Japan, Nos. 23340113, 23104503, and 21740303.
GFB acknowledges support by the 
National Science Foundation under Grant PHY-0835543 and by the  
DOE grant under grant DE-FG02-00ER41132. 
 
\section*{Appendix: Dielectric properties of Sb} 
 
\begin{figure} 
\includegraphics [scale=0.6] {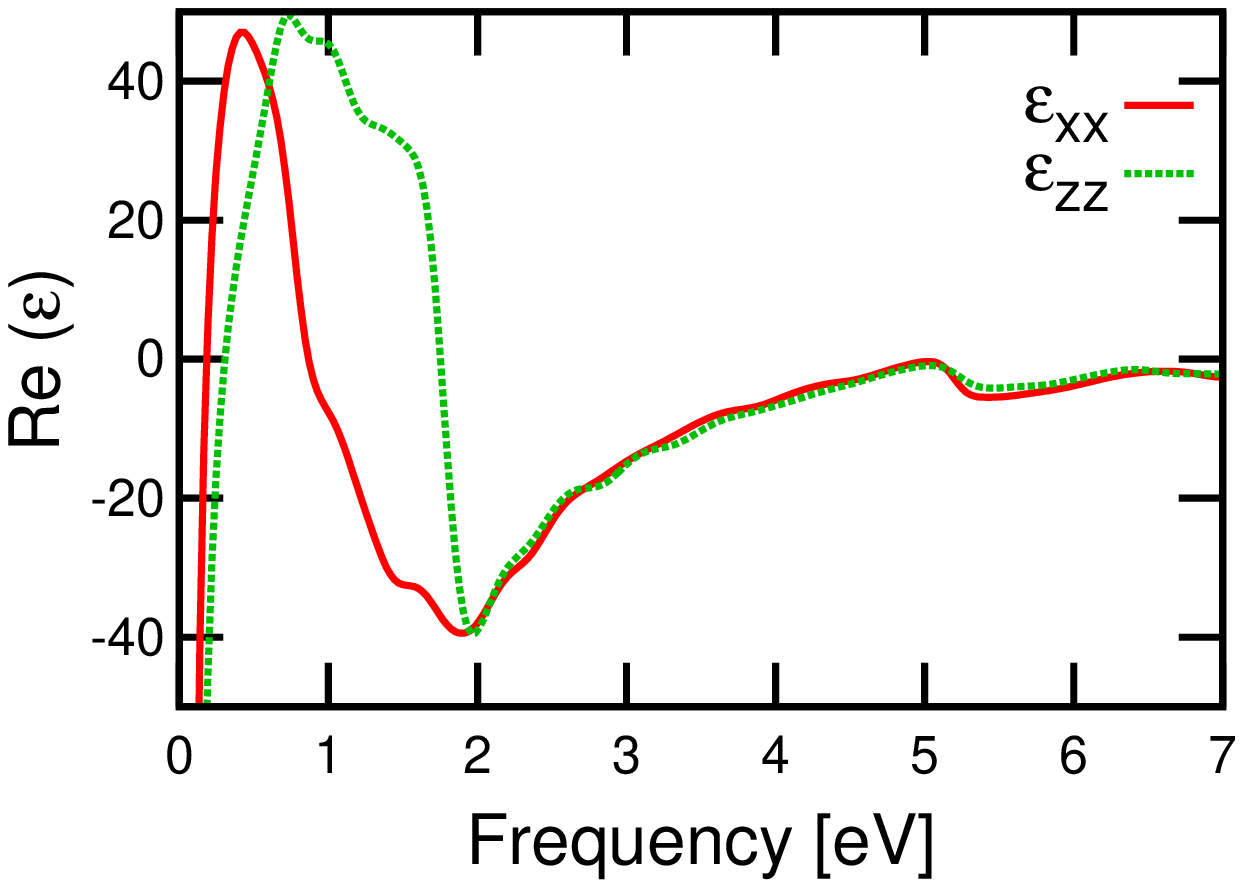} 
\includegraphics [scale=0.6] {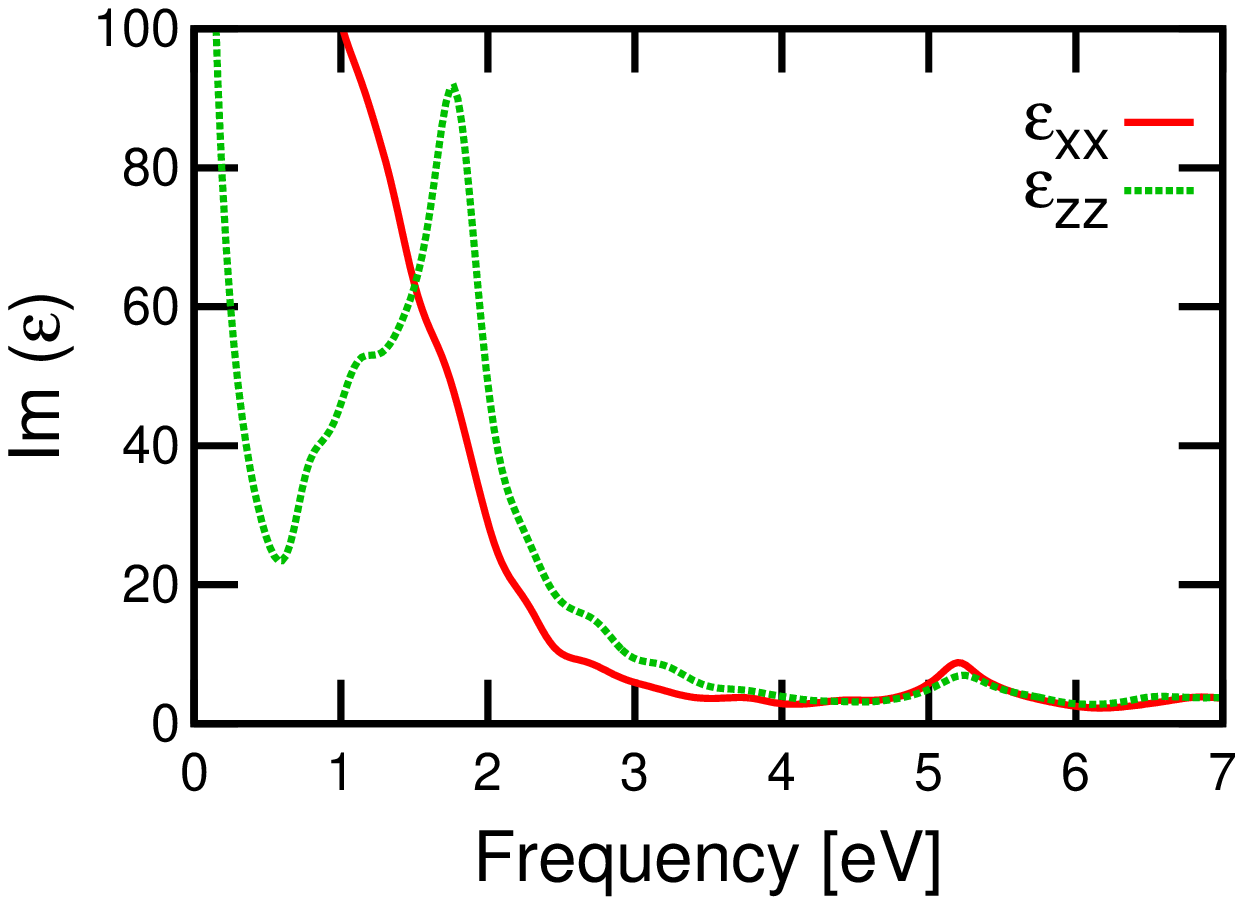} 
\caption{\label{eps_r_i} Calculated dielectric tensor  
$\varepsilon(\omega)$, real-part in the left and imaginary-part 
in the right panels.  Red: $xx$ component; blue: $zz$ component. 
} 
\end{figure}

It is important to check how well the TDDFT performs  
in the linear response region, if one wishes to extend  
the domain to nonlinear processes.   
To that end, we first calculate the dielectric properties  
using the TDDFT and following the real-time method  
proposed in Ref. \cite{biry}.   
There are two independent components of the dielectric tensor  
in crystals of A7 symmetry, $\varepsilon_{xx}$ and $\varepsilon_{zz}$  
in our coordinate system.  They are shown in Fig. \ref{eps_r_i}.   
In the low frequency limit, the real parts are dominated by  
the free carriers and go to negative infinity.   
At intermediate frequencies the real part is very anisotropic  
but becomes isotropic above 2 eV.   
The imaginary part is also very anisotropic in the energy range 
below 3 eV.   
 
\begin{figure} 
\includegraphics [scale=0.6] {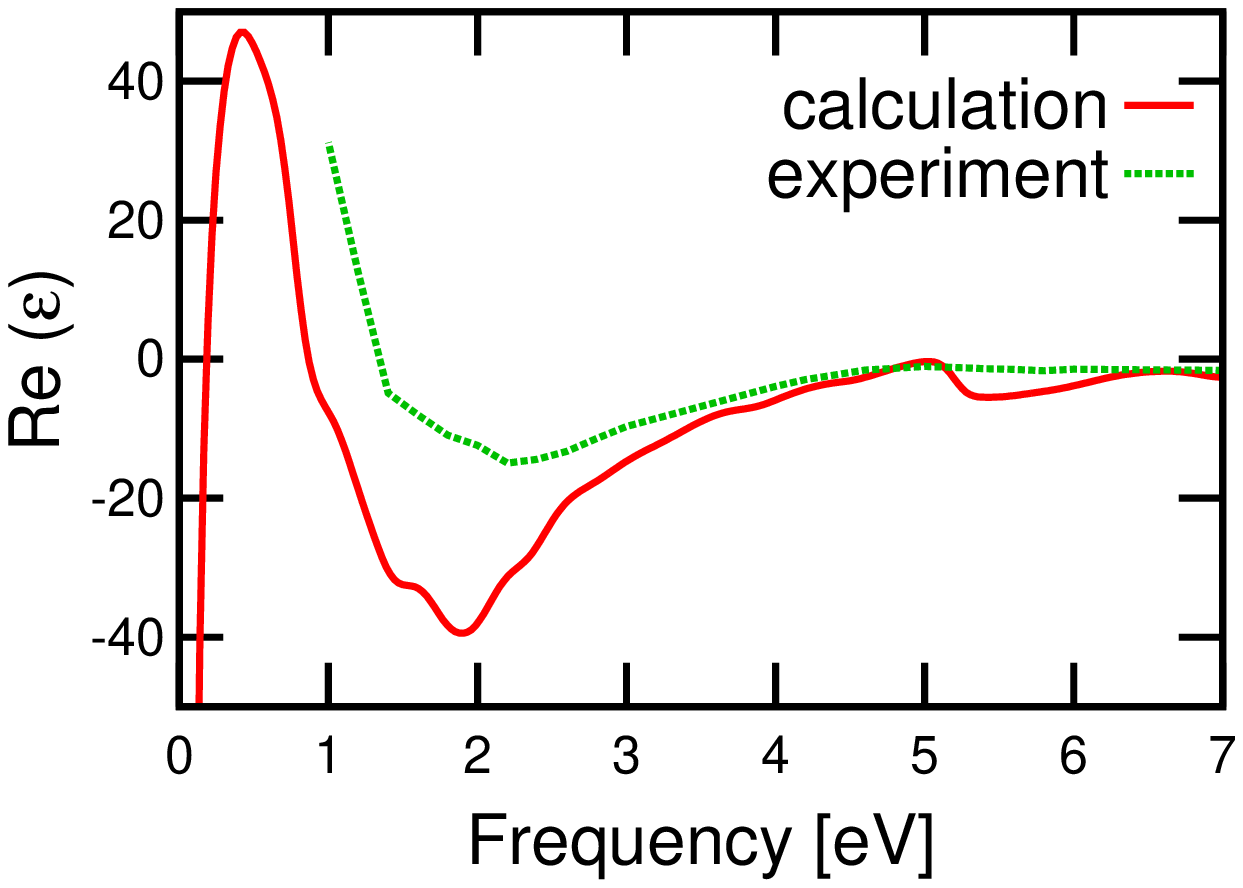} 
\includegraphics [scale=0.6] {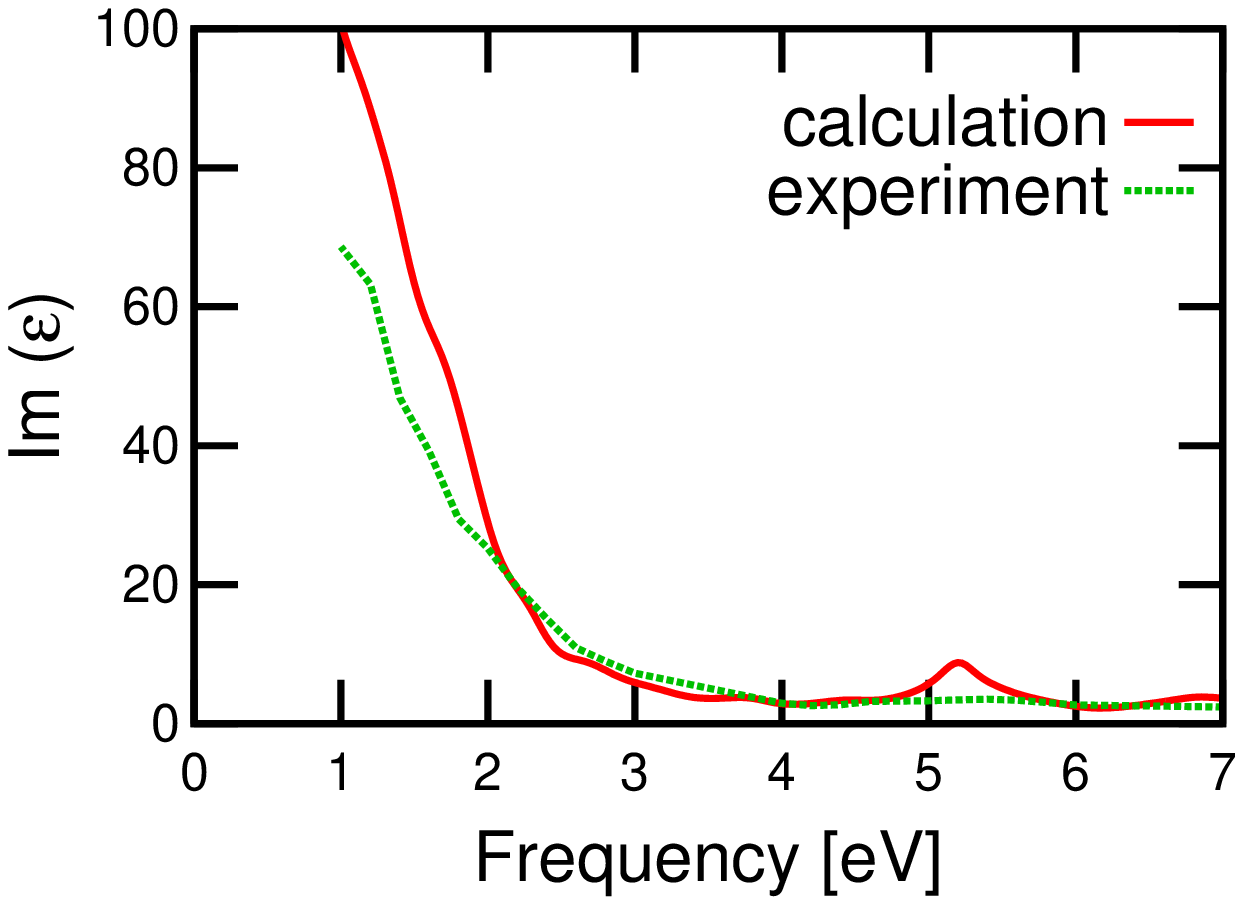} 
\caption{\label{compare-eps} Calculated $\varepsilon_{xx}$ 
compared with evaluated data of Ref. \cite{pa98}. Real-part 
in the left and imaginary-part in the right panels, respectively.  
} 
\end{figure}

The compilation \cite{pa98} of evaluated experimental dielectric functions 
has a table of $\varepsilon_{xx}$ for Sb, which we compare with in 
Fig. \ref{compare-eps}.  In both theory and experiment, one sees large  
absorption strength at low frequencies 
tapering  off smoothly as the frequency increases.  However, the 
agreement is not good at a quantitatively level, unlike the  
situation in simple cubic materials.  The theory is also disappointing 
for describing the real part of $\varepsilon_{xx}$. 
 
\begin{figure} 
\includegraphics [scale=0.6] {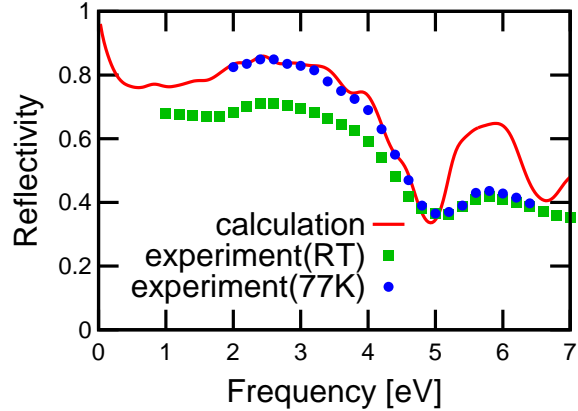} 
\caption{\label{R} Calculated reflectivity of Sb compared with  
experiment \cite{ca64}. 
Line: theory; solid squares:  room temperature data; solid circles:  
data at 77K.  
} 
\end{figure}

In an attempt to get some insight into possible origins of the  
disagreement, we went back to the actual reflectivity data \cite{ca64} 
that were used to obtain the dielectric function.  This data has an 
unexplained temperature dependence with a significantly higher  
reflectivity at 77 K.  The theoretical reflectivity, $R$, is easily  
evaluated as 
\be 
R = \left|{\varepsilon^{1/2} -1\over \varepsilon^{1/2} +1 }\right|. 
\ee 
The comparison between experimental and theoretical $R$ is shown in 
Fig. \ref{R}.  The theory clearly disagrees with the room-temperature 
data.  However, it reproduces rather well the low temperature data 
up to about 5 eV photon energy.  Whether this is a pure accident or 
suggests some missing physics we cannot say. 


\begin{thebibliography}{99}
\bibitem{note1}  While the rapid change around the direct gap is reproduced 
by TDDFT, the energy of the direct gap is typically predicted too 
low in the underlying DFT theory.
\end{thebibliography}
\end{document}